\documentclass[reprint,superscriptaddress,amssymb,aps, pra, twocolumn,graphicx]{revtex4-2}

\usepackage{graphicx}
\usepackage{dcolumn}
\usepackage{bm}
\usepackage{hyperref}

\usepackage{amsmath} 
\usepackage{mathrsfs}
\usepackage[cmintegrals]{newtxmath}
\usepackage[mathlines]{lineno}

\begin{document}


\title{Tunable partial polarization beam splitter and optomechanically induced Faraday effect}

\author{Xuan Mao}
\email{These authors contributed equally to this work}
\affiliation{Department of Physics, State Key Laboratory of Low-Dimensional Quantum Physics, Tsinghua University, Beijing 100084, China}

\author{Guo-Qing Qin}
\email{These authors contributed equally to this work}
\affiliation{Department of Physics, State Key Laboratory of Low-Dimensional Quantum Physics, Tsinghua University, Beijing 100084, China}

\author{Hong Yang}
\author{Zeguo Wang}
\affiliation{Department of Physics, State Key Laboratory of Low-Dimensional Quantum Physics, Tsinghua University, Beijing 100084, China}

\author{Min Wang}
\affiliation{Beijing Academy of Quantum Information Sciences, Beijing 100193, China}

\author{Gui-Qin Li}
\affiliation{Department of Physics, State Key Laboratory of Low-Dimensional Quantum Physics, Tsinghua University, Beijing 100084, China}
\affiliation{Frontier Science Center for Quantum Information, Beijing 100084, China}

\author{Peng Xue}
\email{gnep.eux@gmail.com}
\affiliation{Beijing Computational Science Research Center, Beijing, China}

\author{Gui-Lu Long}
\email{gllong@tsinghua.edu.cn}
\affiliation{Department of Physics, State Key Laboratory of Low-Dimensional Quantum Physics, Tsinghua University, Beijing 100084, China}
\affiliation{Beijing Academy of Quantum Information Sciences, Beijing 100193, China}
\affiliation{Frontier Science Center for Quantum Information, Beijing 100084, China}
\affiliation{Beijing National Research Center for Information Science and Technology, Beijing 100084, China}
\affiliation{School of Information, Tsinghua University, Beijing 100084, China}

\date{\today}

\begin{abstract}

Polarization beam splitter (PBS) is a crucial photonic element to separately extract transverse-electric (TE) and transverse-magnetic (TM) polarizations from the propagating light fields. Here, we propose a concise, continuously tunable and all-optical partial PBS in the vector optomechanical system which contains two orthogonal polarized cavity modes with degenerate frequency. The results show that one can manipulate the polarization states of different output fields by tuning the polarization angle of the pumping field and the system function as partial PBS when the pump laser polarizes vertically or horizontally. As a significant application of the tunable PBS, we propose a scheme of implementing quantum walks in resonator arrays without the aid of other auxiliary systems. Furthermore, we investigate the optomechanically induced Faraday effect in the vector optomechanical system which enables arbitrary tailoring of the input lights and the behaviors of polarization angles of the output fields in the under couple, critical couple, and over couple regimes. Our findings prove the optomechanical system is a potential platform to manipulate the polarization states in multimode resonators and boost the process of applications related to polarization modulation.

\end{abstract}


\maketitle


\section{INTRODUCTION \label{introduction}}

The manipulation of arbitrary polarization states is of significant fundamental and applied relevance to a variety of research fields such as quantum communication networks \cite{xiong2018complete} and quantum optics \cite{nielsen2010quantum}. Polarization beam splitter (PBS) \cite{shen2015integrated, dai2013polarization, li1996visible} plays a significant role in polarization selection. Varies of PBS schemes have been proposed based on Mach-Zehnder interferometer \cite{soldano1994mach}, photonic crystal fiber structure \cite{saitoh2004polarization}, binary blazed grating coupler \cite{feng2007polarization}, multimode interference \cite{hong2003design}, and asymmetrical directional coupler \cite{dai2011ultrashort}. However, previous works barely report a continuously tunable PBS and a general platform that provides tunability of polarization states is needed.

Quantum walks (QW), the quantum correspondence of classical random walks, is proved to be a versatile platform to implement quantum algorithms and simulations\cite{childs2009universal, broome2010discrete, cardano2015quantum, cardano2016statistical, cardano2017detection, childs2013universal, schreiber2011decoherence, crespi2013anderson, xue2015experimental}. QW has been developed in various physical system such as nuclear magnetic resonance \cite{du2003experimental, ryan2005experimental}, coupled waveguides \cite{rai2008transport, tang2018experimental, perets2008realization}, trapped ions \cite{schmitz2009quantum, xue2009quantum}, and photonic systems \cite{peruzzo2010quantum, biggerstaff2016enhancing, cui2020quantum, moqadam2015quantum}. Recently, the QW exhibits various topological phases \cite{fu2007topological, schnyder2008classification, qi2008topological, rechtsman2013photonic} and demonstrates fascinating topological phenomena \cite{susstrunk2015observation, xiao2017observation, xiao2021observation, xiao2020non}. However, there is short of schemes that using the internal degrees of freedom as coin states in resonator arrays without the aid of other auxiliary systems.

High-quality whispering gallery mode (WGM) microcavities \cite{vahala2003optical} have potential value in investigating fundamental physics and practical technologies such as cavity optomechanics \cite{aspelmeyer2014cavity, weis2010optomechanically, kronwald2013optomechanically, dong2012optomechanical, jiang2015chip, wang2019characterization, safavi2011electromagnetically, fiore2011storing, qin2020manipulation, liao2016macroscopic, lu2015p, shen2016experimental}, low-threshold lasing \cite{feng2014single, hu2021demonstration, jing2014pt, lu2017exceptional, zhang2018phonon, spillane2002ultralow}, quantum sensing \cite{degen2017quantum, chen2017exceptional, liang2021low, qin2021experimental, djorwe2019exceptional, lai2020earth, mao2020enhanced, qin2019brillouin, khial2018nanophotonic}, and nonlinear optics \cite{fan2012all, kronwald2013optomechanically, kippenberg2004kerr, zhang2019symmetry, zhu2019controllable} due to their ability to enhance light-matter interactions. Characterized by exploring the radiation pressure interaction between optical modes and mechanical modes, optomechanics exhibits rich physical phenomena such as optomechanically induced transparency (OMIT) \cite{weis2010optomechanically, qin2020manipulation, kronwald2013optomechanically, kim2015non}, absorption (OMIA) \cite{qin2020manipulation, naweed2005induced, qu2013phonon}, and optomechanically induced Faraday effect (OMIFE) \cite{duggan2019optomechanically}. These effects enable a new degree of light control and achieve arbitrary tailoring of the input lights in optomechanical systems. Further, the additional degree of light control allows varies of applications including state transfer \cite{tian2012adiabatic, tian2010optical, xu2020frequency, zhang2019fast, wang2012using}, optical routing \cite{ruesink2018optical, yang2020multimode, shen2018reconfigurable}, and entanglement generation \cite{wang2013reservoir, kuzyk2013generating, wang2015bipartite}. Besides progressing in many applications such as frequency comb generation \cite{del2007optical, domeneguetti2021parametric} and light storage \cite{fiore2011storing, dong2015brillouin}, optoemchanical systems provides a promising platform to study polarization behaviors.

In this paper, we theoretically propose a concise, continuously tunable and all-optical partial PBS in the vector optomechanical system which contains two optical modes coupling with the same mechanical mode. Since the effective refractive indexes are polarization depend in the resonators, the two optical modes with orthogonal polarizations and degenerate frequency can be achieved. In this content, we consider the pump and probe fields are both linearly polarized. We study the transmission spectra of different ports with different polarizations. Specifically, when the included angle between the polarization vector of the pump field and the horizontal mode equals to 0, the output field of port 2 polarizes vertically only  while the polarization of the output field of port 4 is parallel to the horizontal mode in the case of resonance. Thus, the vector optomechanical system function as PBS and it turns over the result when tuning the polarization of pump field from horizontal to vertical. As a significant application of the tunable PBS, we propose a scheme of implementing QW in resonator arrays without the aid of other auxiliary systems. Furthermore, OMIFE enables arbitrary tailoring of the input fields in the system and we investigate the polarization behaviors of the output fields in the under couple, critical couple, and over couple regimes. We believe that our findings evidence the optomechanical system is a potential platform to manipulate the polarization states in multimode resonators and boost the process of applications related to polarization modulation.

This article is organized as follows: In Sec.\ref{basic model}, we demonstrate the basic model and the dynamical equations. We study the transmission spectra in Sec.\ref{transmission}. We show the OMIFE in Sec.\ref{faraday effect}. Conclusion is given in Sec.\ref{conclusion}.

\section{MODEL AND DYNAMICAL EQUATIONS\label{basic model}}

\begin{figure}
    \centering
    \includegraphics[width=\linewidth]{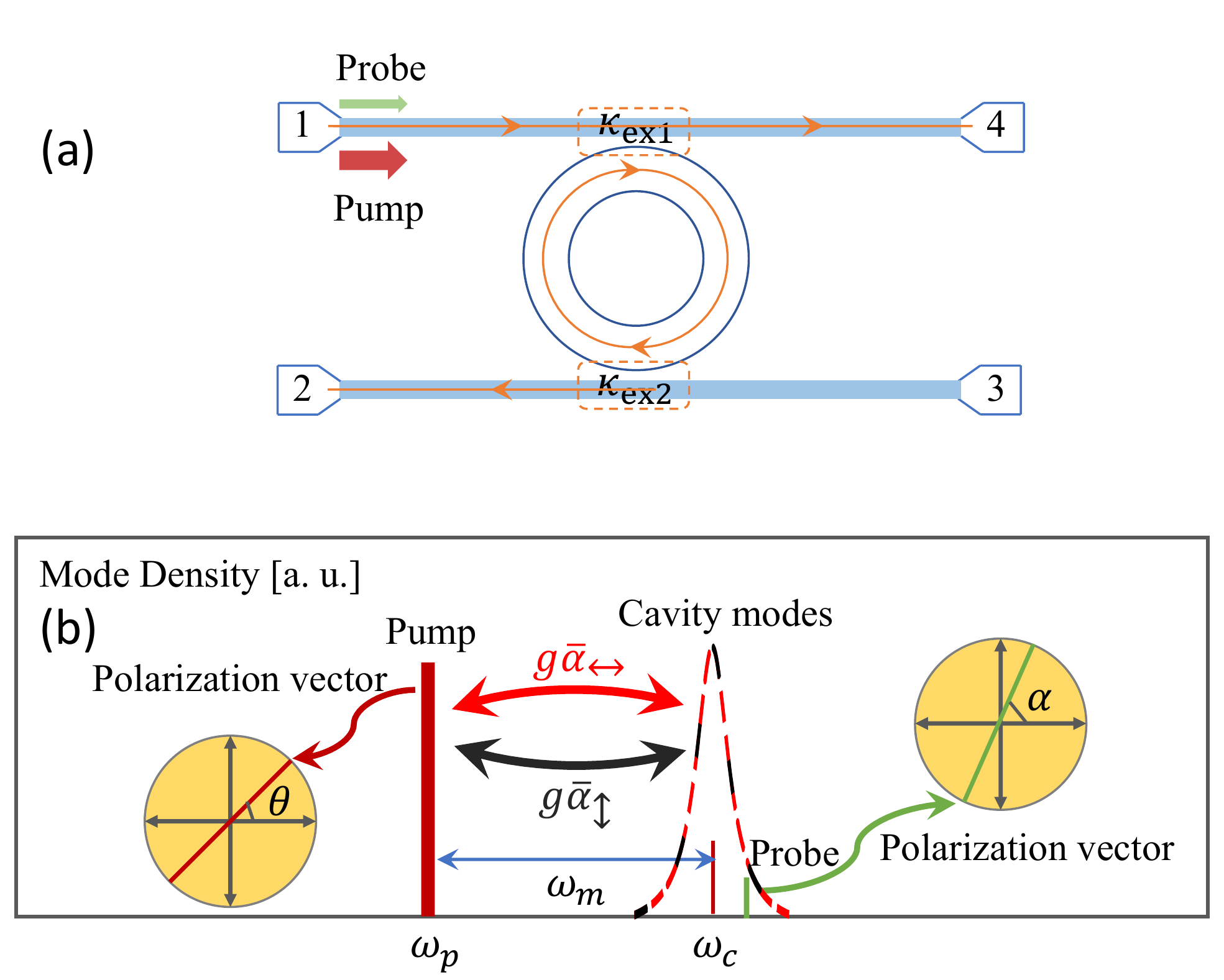}
    \caption{(a) Schematic of the vector optomechanical system. (b) Frequency spectrogram of the vector optomechanical system, which is composed of two degenerate cavity modes with orthogonal polarizations. The pump and probe fields are both linearly polarized and the included angle between the polarization vector of the pump (probe) field and the horizontal mode is $\theta$ ($\alpha$).}
    \label{model}
\end{figure}

The vector optomechanical model we proposed is illustrated in Fig. \ref{model}  which contains two degenerate optical modes, with degenerate frequency $\omega_c$ and decay rate $\kappa$, coupling with the same mechanical mode characterized by frequency $\omega_m$ and the damping constant $\Gamma_m$. The Hamiltonian of our system pumped by the linearly optical field can be described by ($\hbar = 1$)

\begin{align}
    H &= H_{free} + H_{int} + H_{drive} + H_{probe}, \label{equation 1}
\end{align}

where 

\begin{align}
    H_{free} =& \omega_{c} (a_{\updownarrow}^\dagger a_{\updownarrow} + a_{\leftrightarrow}^\dagger a_{\leftrightarrow}) + \omega_{m} b^\dagger b, \nonumber \\
    H_{int} =& g a_{\updownarrow}^\dagger a_{\updownarrow} (b^\dagger + b) + g a_{\leftrightarrow}^\dagger a_{\leftrightarrow} (b^\dagger + b), \nonumber\\
    H_{drive} =& i \epsilon_{p\updownarrow} \sqrt{\kappa_{ex1}} e^{-i \omega_p t} a_{\updownarrow}^\dagger  + i \epsilon_{p\leftrightarrow} \sqrt{\kappa_{ex1}} e^{-i \omega_p t} a_{\leftrightarrow}^\dagger + H.c.,\nonumber\\
    H_{probe} =& i \epsilon_{r\updownarrow} \sqrt{\kappa_{ex1}} e^{-i \omega_r t} a_{\updownarrow}^\dagger + i \epsilon_{r\leftrightarrow} \sqrt{\kappa_{ex1}} e^{-i \omega_r t} a_{\leftrightarrow}^\dagger + H.c. , \label{equation 2}
\end{align}

$H_{free}$ describes the free Hamiltonian of the optomechanical system, $a_{j}$ and $a_j^\dagger$ (for $j=\updownarrow,\leftrightarrow$) are the annihilation and creation operators of the optical mode, $\updownarrow$ and $\leftrightarrow$ label the vertical and horizontal polarization axes respectively. 
The mechanical annihilation and creation operators are denoted by $b$ and $b^\dagger$. $H_{int}$ characterizes the interaction Hamiltonian of the system with the single-photon optomechanical coupling strength $g$. $H_{drive}$ implies the two degenerate optical modes are driven by external fields with strength $\epsilon_{pj}$ and frequency $\omega_{p}$. As illustrated in Fig. \ref{model}(b), $\theta$ describes the included angle between the polarization vector of the driving field and the horizontal axes. Hence $\epsilon_{p\updownarrow} = \epsilon_{p} \sin(\theta)$, $\epsilon_{p\leftrightarrow} = \epsilon_{p} \cos(\theta)$ where $\epsilon_p = \sqrt{P_{in}/(\hbar \omega_p)}$ and $P_{in}$ is the input power of the driving field. $\kappa_{ex1}$ denotes the external loss rate between the optical mode $a_j$ and the fiber. $H_{probe}$ describes the probe laser characterized by strength $\epsilon_{rj}$ and frequency $\omega_r$. $\epsilon_{r\updownarrow}$ and $\epsilon_{r\leftrightarrow}$ satisfy $\epsilon_{r\updownarrow}/\epsilon_{r\leftrightarrow} = \tan(\alpha)$ where $\alpha$ denotes the angle between the polarization vector of the probe laser and the horizontal axes. $\epsilon_r = \sqrt{P_r/(\hbar \omega_r)}$ and $P_r$ denotes the input power of the probe field. In the rotating frame with the driving fields and after following the standard linearization procedure, the linearized equations of the fluctuation parts are expressed as

\begin{align}
    \frac{\mathrm{d} a_{\updownarrow}}{\mathrm{d} t} &= -(i \Delta + \frac{\kappa}{2}) a_{\updownarrow} - i G_{\updownarrow} b + \sqrt{\kappa_{ex1}} \epsilon_{r\updownarrow} e^{-i \delta t},  \label{equation 3}\\
    \frac{\mathrm{d} a_{\leftrightarrow}}{\mathrm{d} t} &= -(i \Delta + \frac{\kappa}{2}) a_{\leftrightarrow} - i G_{\leftrightarrow} b + \sqrt{\kappa_{ex1}} \epsilon_{r\leftrightarrow} e^{-i \delta t},  \label{equation 4}\\
    \frac{\mathrm{d} b}{\mathrm{d} t} &= -(i \omega_{m} + \frac{\Gamma_m}{2}) b - i G_{\updownarrow} a_{\updownarrow} - i G_{\leftrightarrow} a_{\leftrightarrow}.  \label{equation 5}
\end{align}

Here, $\Delta = \omega_{c} - \omega_p$ represents the detuning between the optical mode and the driving field. $\delta = \omega_r - \omega_p$ is the detuning between the probe laser and the control field. $G_{\updownarrow}$ ($G_{\leftrightarrow}$) is the effective optomechanical coupling strength between the vertical(horizontal) optical mode and the mechanical mode. The solutions of the Eq. \ref{equation 3}-Eq. \ref{equation 5} are given by 

\begin{align}
    a_{\updownarrow} &= \frac{\sqrt{\kappa_{ex1}} \epsilon_{r\updownarrow} - i G_{\updownarrow} b}{\beta_1},  \label{equation 6}\\
    a_{\leftrightarrow} &= \frac{\sqrt{\kappa_{ex1}} \epsilon_{r\leftrightarrow} - i G_{\leftrightarrow} b}{\beta_1},  \label{equation 7}\\
    b &= - \frac{i \sqrt{\kappa_{ex1}}(G_{\updownarrow} \epsilon_{r\updownarrow} + G_{\leftrightarrow} \epsilon_{r\leftrightarrow})}{\beta_m \beta_1 + G_{\updownarrow}^2 + G_{\leftrightarrow}^2},  \label{equation 8}
\end{align}

$\beta_1 = i \Delta + \kappa/2$ and $\beta_m = i \omega_{m} + \Gamma_m/2$. The output fields of the optomechanical system can be obtained by adopting the input-output relation $\epsilon_{out} = \epsilon_{in} - \sqrt{\kappa_{ex}} a$. Specifically, the output field of port 2 and port 4 are expressed as  

\begin{align}
    \vec{\epsilon}_{2out} &= -\sqrt{\kappa_{ex2}} a_{\updownarrow} \vec{e}_{\updownarrow} -\sqrt{\kappa_{ex2}} a_{\leftrightarrow} \vec{e}_{\leftrightarrow}, \label{equation 9}\\
    \vec{\epsilon}_{4out} &= (\epsilon_{r\updownarrow} -\sqrt{\kappa_{ex1}} a_{\updownarrow}) \vec{e}_{\updownarrow} + ( \epsilon_{r\leftrightarrow} -\sqrt{\kappa_{ex1}} a_{\leftrightarrow}) \vec{e}_{\leftrightarrow}. \label{equation 10}
\end{align}

Here, $\vec{e}_{\updownarrow}$ and $\vec{e}_{\leftrightarrow}$ are the unit vectors of the vertical mode and the horizontal mode, respectively.

\section{TRANSMISSION SPECTRA ANALYSIS \label{transmission}}

\begin{figure*}
    \centering
    \includegraphics[width=\linewidth]{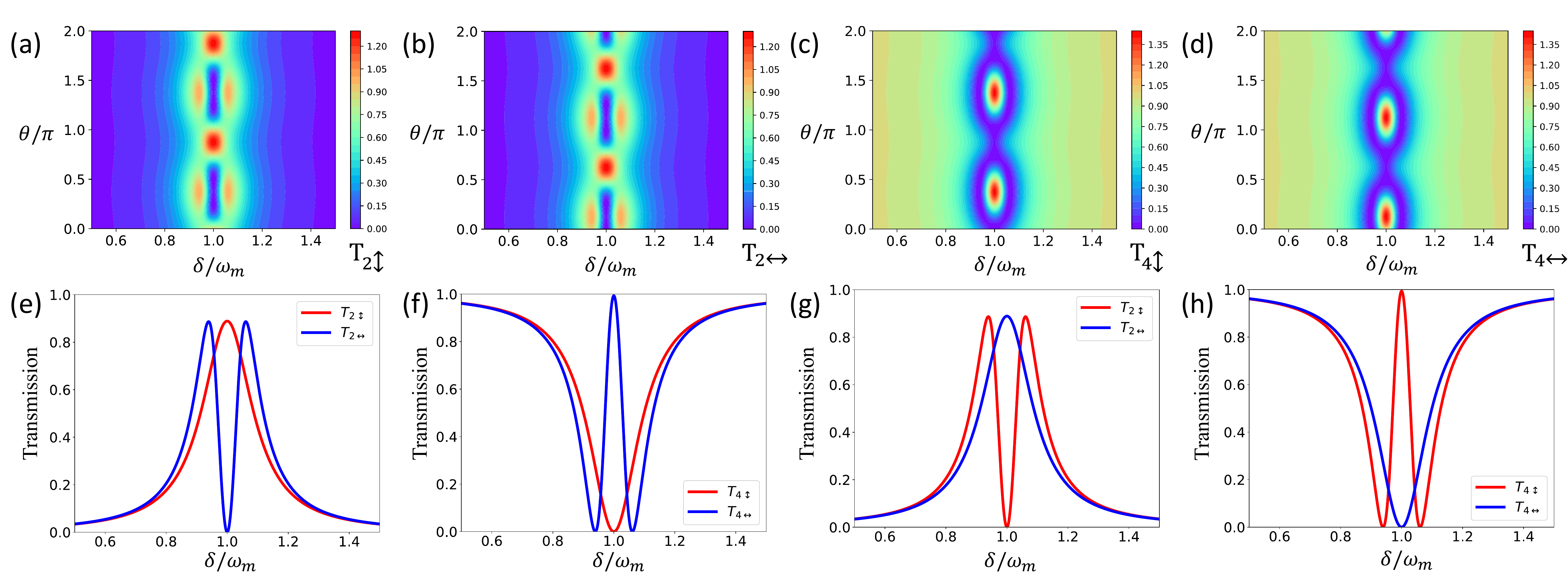}
    \caption{Transmissions of different ports with different polarizations as a function of $\theta/\pi$ and $\delta/\omega_m$: (a) port 2 with vertical polarization, (b) port 2 with horizontal polarization, (c) port 4 with vertical polarization, and (d) port 4 with horizontal polarization. (e) and (f) illustrate the transmissions of port 2 and port 4 when $\theta = 0$. (g) and (h) illustrate the transmissions of port 2 and port 4 when $\theta = \pi/2$. The parameters used in this system are $\kappa_{0\updownarrow}/2\pi = \kappa_{0\leftrightarrow}/2\pi = \kappa_{0} / 2\pi = 1$ MHz, $\kappa_{ex1}/2\pi = 9$ MHz, $\kappa_{ex2}/2\pi = 8$ MHz, $\omega_m/2\pi = 90.47$ MHz, $\Gamma_m/2\pi = 22$ kHz, $G_{\updownarrow}/2\pi = 5.5 \times sin(\theta)$ MHz, $G_{\leftrightarrow}/2\pi = 5.5 \times cos(\theta)$ MHz, $c = 3 \times 10^8$ m/s, $\lambda = 1550$ nm, $P_r = 20 \mu$W, and $\alpha = \pi/4$.}
    \label{PBS}
\end{figure*}

In Eq. \ref{equation 9} and Eq. \ref{equation 10}, higher-order sidebands are not considered, one can obtain the normalized transmission coefficients of different polarizations out of port 2 and port 4, i.e., $tran_{2\updownarrow}$, $tran_{2\leftrightarrow}$, $tran_{4\updownarrow}$, and $tran_{4\leftrightarrow}$. The normalized transmission coefficients link input to output modes,

\begin{align}
    \left[
    \begin{array}{c}
        \epsilon_{2out\updownarrow}\\
        \epsilon_{2out\leftrightarrow}\\
        \epsilon_{4out\updownarrow}\\
        \epsilon_{4out\leftrightarrow}
    \end{array}
    \right]
    =
    \left[
    \begin{array}{ccccc}
        tran_{2\updownarrow} & 0 & 0 & 0\\
        0 & tran_{2\leftrightarrow} & 0 & 0\\
        0 & 0 & tran_{4\updownarrow} & 0\\
        0 & 0 & 0 & tran_{4\leftrightarrow}
    \end{array}
    \right]
    \times
    \left[
    \begin{array}{c}
        \epsilon_{r\updownarrow}\\
        \epsilon_{r\leftrightarrow}\\
        \epsilon_{r\updownarrow}\\
        \epsilon_{r\leftrightarrow}
    \end{array}
    \right] \nonumber\\
    =
    \left[
    \begin{array}{ccccc}
        -\frac{\sqrt{\kappa_{ex2}} a_{\updownarrow}}{\epsilon_{r\updownarrow}} & 0 & 0 & 0\\
        0 & -\frac{\sqrt{\kappa_{ex2}} a_{\leftrightarrow}}{\epsilon_{r\leftrightarrow}} & 0 & 0\\
        0 & 0 & 1 - \frac{\sqrt{\kappa_{ex1}} a_{\updownarrow}}{\epsilon_{r\updownarrow}} & 0\\
        0 & 0 & 0 & 1 - \frac{\sqrt{\kappa_{ex1}} a_{\leftrightarrow}}{\epsilon_{r\leftrightarrow}}
    \end{array}
    \right]
    \times
    \left[
    \begin{array}{c}
        \epsilon_{r\updownarrow}\\
        \epsilon_{r\leftrightarrow}\\
        \epsilon_{r\updownarrow}\\
        \epsilon_{r\leftrightarrow}
    \end{array}
    \right], \label{equation 11}
\end{align}
where $\epsilon_{2out\updownarrow}$ ($\epsilon_{4out\updownarrow}$) and $\epsilon_{2out\leftrightarrow}$ ($\epsilon_{4out\leftrightarrow}$) are the projections of $\vec{\epsilon}_{2out}$ ($\vec{\epsilon}_{4out}$) onto the vertical and horizontal modes, respectively. Furthermore, the transmission rate is the square of the corresponding normalized transmission coefficient. For instance, the transmission rate of the vertical field out of port 2 is $T_{2\updownarrow} = |tran_{2\updownarrow}|^2$. 

As analysis above, the transmission of different ports with different polarizations can be tuned by changing the related parameter values. The results show that the vector optomechanical system function as tunable PBS for some specific parameters. Considering the experimental feasibility \cite{shen2018reconfigurable}, the parameters used in this system are the intrinsic decay rate of the two optical modes $\kappa_{0\updownarrow}/2\pi = \kappa_{0\leftrightarrow}/2\pi = \kappa_{0} / 2\pi = 1$ MHz, $\kappa_{ex1}/2\pi = 9$ MHz, $\kappa_{ex2}/2\pi = 8$ MHz, $\omega_m/2\pi = 90.47$ MHz, $\Gamma_m/2\pi = 22$ kHz, $G_{\updownarrow}/2\pi = 5.5 \times sin(\theta)$ MHz, $G_{\leftrightarrow}/2\pi = 5.5 \times cos(\theta)$ MHz, $c = 3 \times 10^8$ m/s, $\lambda = 1550$ nm, the power of the probe field $P_r = 20 \mu$W, and $\alpha = \pi/4$. 

The transmission spectra of port 2 and port 4 with different polarizations are demonstrated in Fig. \ref{PBS}. Fig. \ref{PBS} (a) - Fig. \ref{PBS} (d) illustrate $T_{2\updownarrow}$, $T_{2\leftrightarrow}$, $T_{4\updownarrow}$, and $T_{4\leftrightarrow}$ as a function of the included polarization angle of the pump laser $\theta$ in the unit of $\pi$ and the detuning $\delta$ in the unit of $\omega_m$, respectively. It is clear that the transmission rate changes periodically with $\theta$ and the period is $\pi$ no matter which port the output belongs to or which polarization the output field is. Note that the transmission rate in Fig. \ref{PBS} exceeds 1 in some regions which never happens in regular transmission spectra. The physical interpretation is there is Faraday effect induced by optomechanics in the vector system. The polarization of the incident probe laser experiences rotation related to the polarization angle of the pump field. The details about optomechanical induced Faraday effect(OMIFE) can be found in Sec. \ref{faraday effect}. 

It is interesting that the optomechanical vector system can function as tunable PBS when $\theta$ equals to some specific values. Fig. \ref{PBS} (e) and Fig. \ref{PBS} (f) demonstrate the transmission spectra of port 2 and port 4 when $\theta = 0$. If the detuning between the pump laser and the probe field $\delta$ equals to the mechanical frequency $\omega_m$, the output field of port 2 polarizes vertically only and has no projection onto the horizontal mode. On the contrary, the polarization of the output field of port 4 is parallel to the horizontal mode. The physics behind the phenomenon is that when $\theta = 0$ there is driving field only for the horizontal mode $a_{\leftrightarrow}$. Due to the interference effect between two pathways, optomechanical induced transparence(OMIT) emerges for the horizontal mode. The first pathway is the probe photons excite optical mode $a_{\leftrightarrow}$ and couple to the output port 4 and the other one is the photons generated by the sideband transition through the optomechanical interaction are coupled out of the cavity. For the vertical mode $a_{\updownarrow}$, there is no driving field and the OMIT cannot be observed. As expected, the transmission rate $T_{4\updownarrow}$ exhibits a Lorenz curve. In parallel, Fig. \ref{PBS} (g) and Fig. \ref{PBS} (h) show the PBS can turn over the result when tuning the value of $\theta$ to $\pi/2$. 

\begin{figure*}
    \centering
    \includegraphics[width=0.9\linewidth]{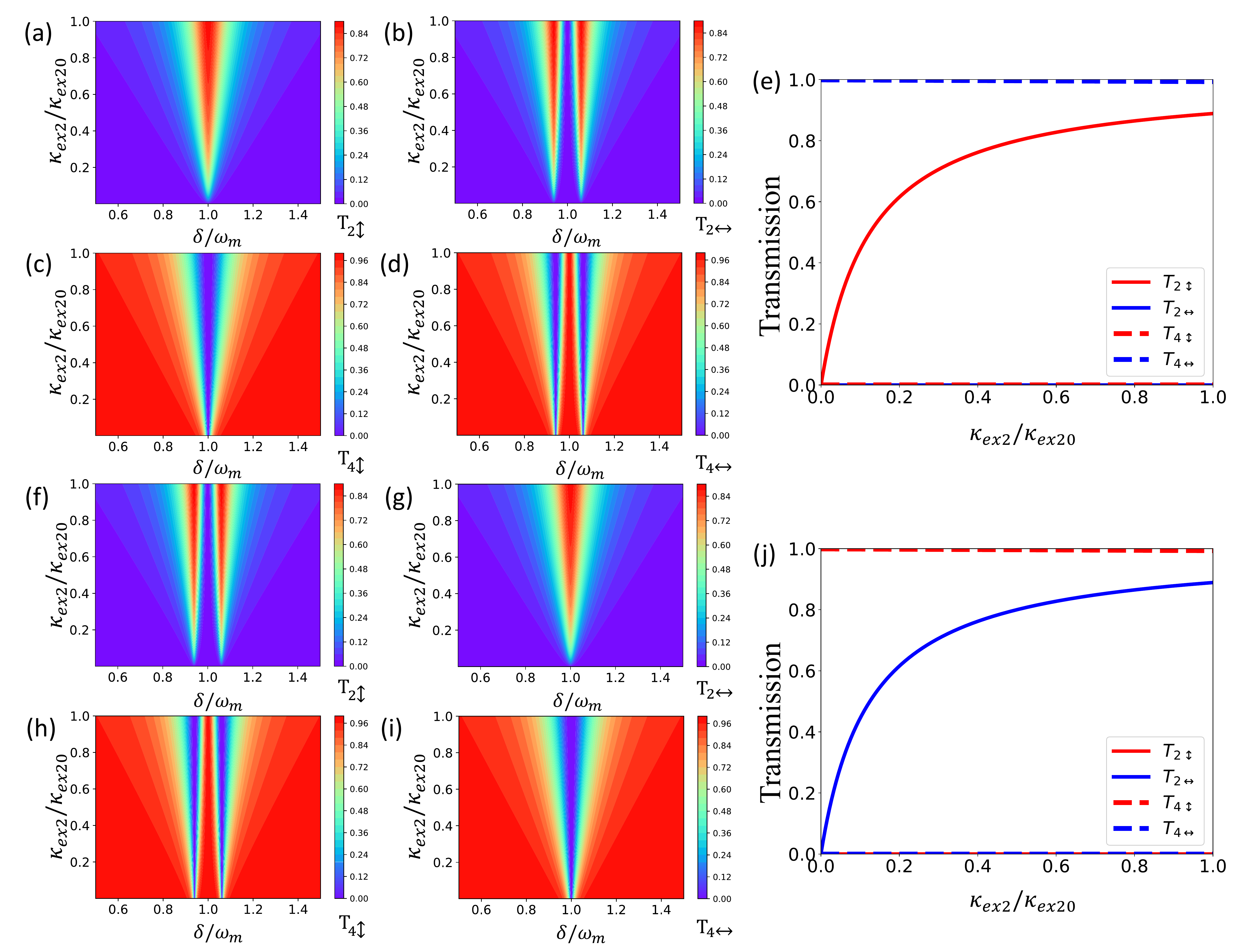}
    \caption{The transmission of different ports with different polarizations as a function of $\kappa_{ex2}/\kappa_{ex20}$ and $\delta/\omega_m$. (a)-(e) are in the case of $\theta = 0$ and (f)-(j) are in the case of $\theta = \pi/2$. (e) and (j) indicate the transmission rate of different ports with vertical or horizontal polarization when tuning the value of $\kappa_{ex2}$ in the case of resonance $\delta = \omega_m$. $\kappa_{ex20}/2\pi = 8$ MHz and $\kappa_{ex1} = \kappa_{ex2} + \kappa_0$.
    The other parameters are the same as that in Fig. \ref{PBS}.}
    \label{kappa}
\end{figure*}

Note that the transmission of port 2 with vertical polarization in the case of $\delta = \omega_m$ and $\theta = 0$ is not 1 due to the presence of the loss in the system. For practical applications, the loss of different polarization states is also should be manipulated to meet experimental requirements. We investigate the impact of the coupling rate $\kappa_{ex2}$ on the transmission rates as shown in Fig. \ref{kappa}. To make sure the polarization of port 2 and port 4 is either vertical or horizontal only, the critical couple condition $\kappa_{ex1} = \kappa_{ex2} + \kappa_{0}$ should be maintained. Take $\theta = 0$ for example, the results show the value of $\kappa_{ex2}$ has a big impact on the linewidth of the output fields of port 2 and port 4, which has been demonstrated by Fig. \ref{kappa}(a) and Fig. \ref{kappa}(c). Under the condition of not changing the polarization state of the output of port 2 and port 4, the transmission rate of the resonance of $\delta = \omega_m$ varies from 0 to 0.9 by tuning the value of $\kappa_{ex2}$ which can be realized by changing the distance between the below fiber and the resonator are shown in Fig. \ref{model}(a). Fig. \ref{kappa}(e) plots the transmission rate of port 2 and port 4 with different polarization in the resonance of $\delta = \omega_m$. It is evident that the transmission of the vertical polarization field of port 2 can be adjusted and so does the loss while the loss of the other polarization state of port 2 and port 4 is maintained. Correspondingly, the loss of the horizontal polarization field of port 4 can also be tuned in the case of $\theta = \pi/2$ and $\delta = \omega_m$. 

As mentioned above, the transmission rate in Fig. \ref{PBS} may exceed 1 in some regions while the total transmission rate of port 2 or port 4 will not. Unlike Eq. \ref{equation 11}, the transmission rates of port 2 and port 4 are given by

\begin{align}
    T_2 &= \frac{|- \sqrt{\kappa_{ex2}} a_{\updownarrow}|^2 + |- \sqrt{\kappa_{ex2}} a_{\leftrightarrow}|^2}{|\epsilon_r|^2}, \label{equation 12}\\
    T_4 &= \frac{|\epsilon_{r\updownarrow} - \sqrt{\kappa_{ex1}} a_{\updownarrow}|^2 + |\epsilon_{r\leftrightarrow} - \sqrt{\kappa_{ex1}} a_{\leftrightarrow}|^2}{|\epsilon_r|^2}. \label{equation 13}
\end{align}

\begin{figure}
    \centering
    \includegraphics[width=\linewidth]{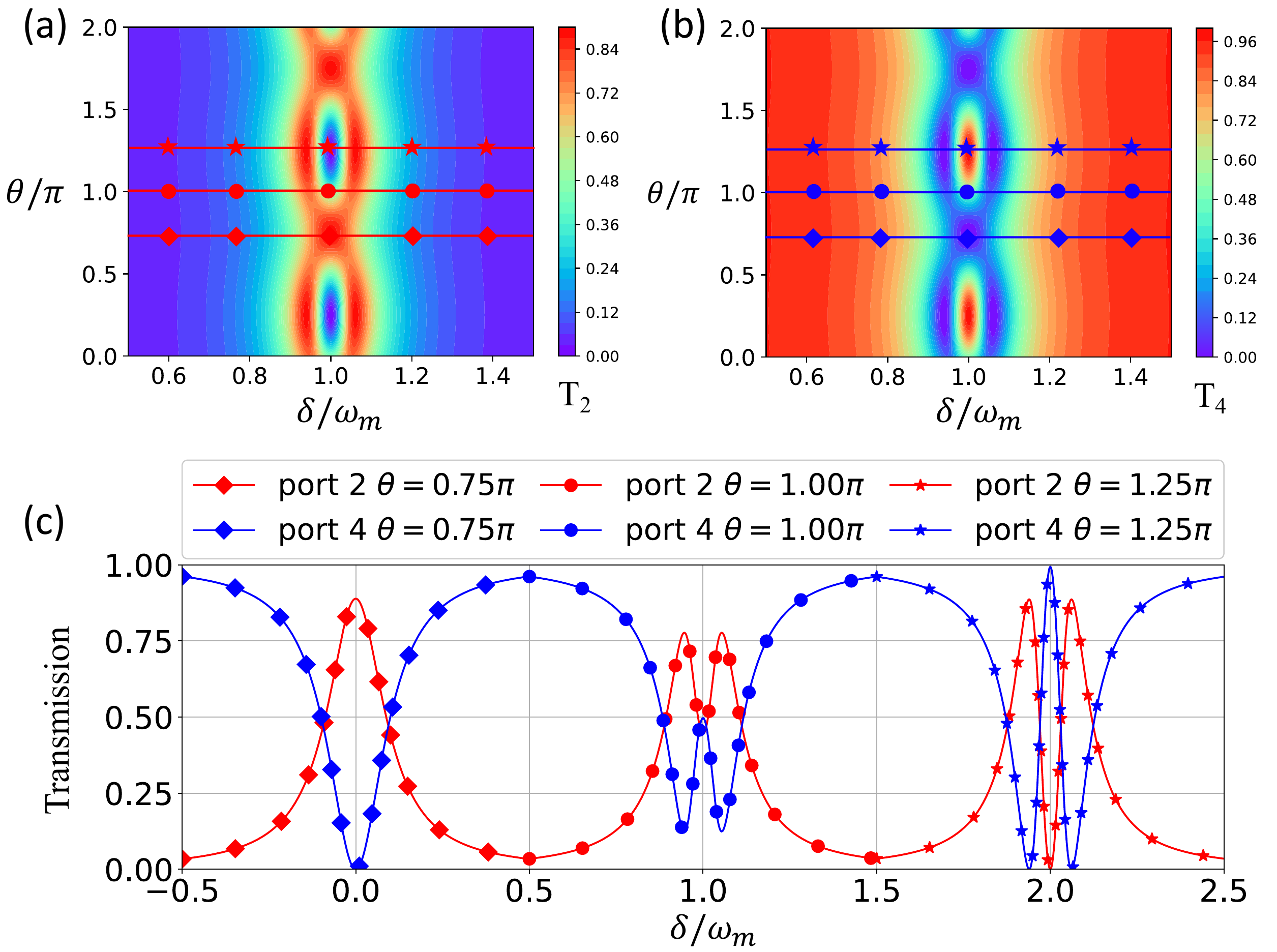}
    \caption{The total transmission rate as a function of $\theta / \pi$ and $\delta / \omega_m$: (a) port 2 $T_2$ and (b) port 4 $T_4$. (c) The total transmission rate $T_2$ and $T_4$ as a function of $\delta / \omega_m$ with different value of $\theta$. To make it clear, note that the transmission rates when $\theta = 0.75 \pi$ and $\theta = 1.25 \pi$ have -$\omega_m$ and $\omega_m$ shift in the $\delta$ axis, respectively. The other parameters are the same as that in Fig. \ref{PBS}.}
    \label{total}
\end{figure}

Fig. \ref{total} shows the total transmission rate of port 2 and port 4 as a function of $\theta / \pi$ and $\delta / \omega_m$. Similar to Fig. \ref{PBS} (a) - Fig. \ref{PBS} (d), the total transmission rates feature a period with $\pi$ as $\theta$ increases. In the domain of Fig. \ref{total} (a) and Fig. \ref{total} (b), the total transmission rates $T_2$ and $T_4$ are always between 0 and 1. Fig. \ref{total} (c) presents the transmission spectra with different value of $\theta$ and are marked by different markers. Note that the markers are corresponding to the markers in Fig. \ref{total} (a) and Fig. \ref{total} (b) according to the value of $\theta$. Further, we have shift the transmission spectra when $\theta = 0.75\pi$ and $\theta = 1.25\pi$ with amount of -$\omega_m$ and $\omega_m$ in the axis of $\delta$ to make it clear. For port 4, the transmission spectrum varies from a typical Lorenz curve to OMIT as $\theta$ changes from 0.75 $\pi$ to 1.25 $\pi$. It can be inferred that the angle of polarization of pump field has a big impact on the optomechanical interference effect and influences the transmission spectrum further. 

\begin{figure}
    \centering
    \includegraphics[width=\linewidth]{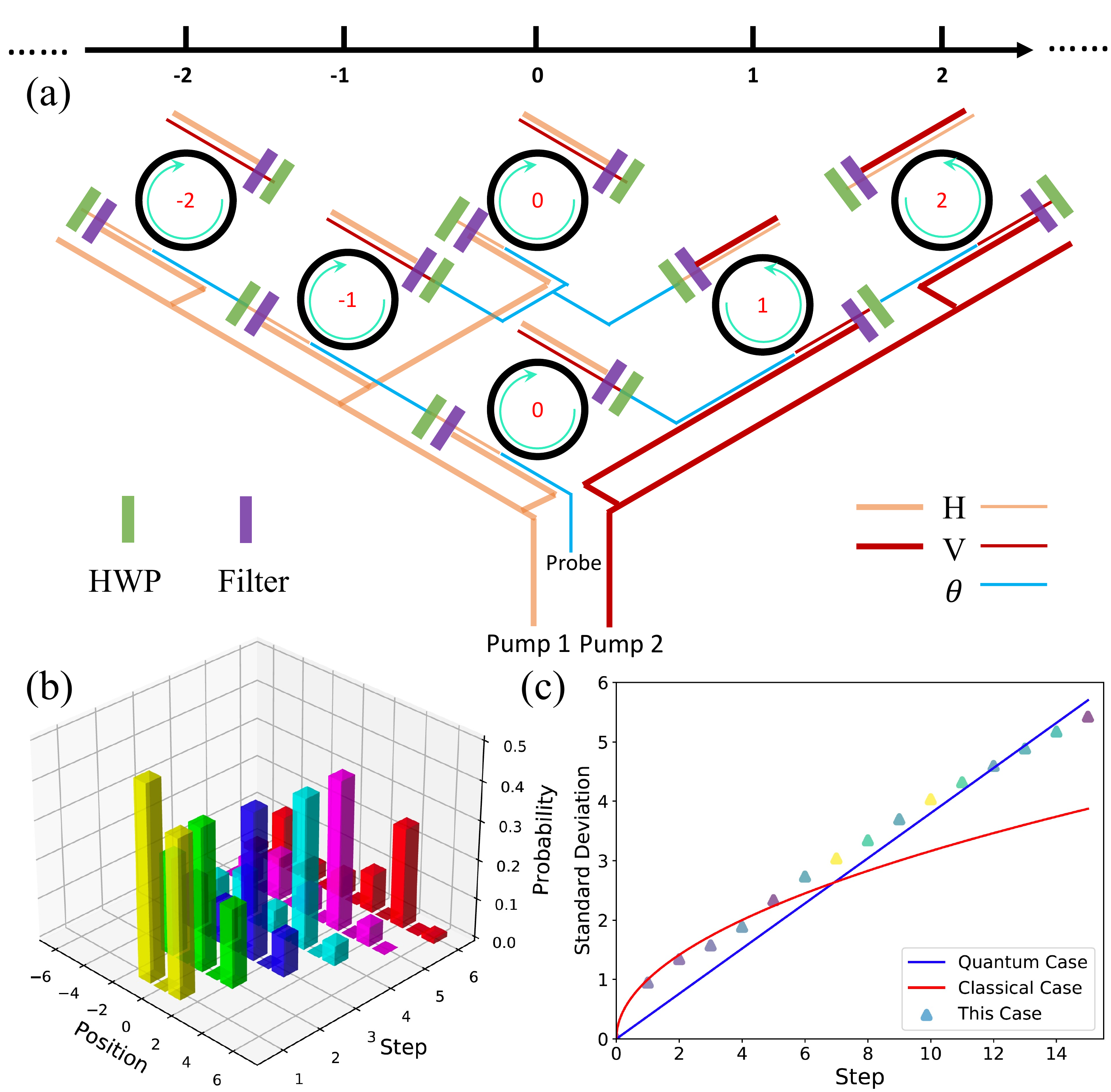}
    \caption{(a) Schematic of implementation of QW in optomechanical system. HWP: half wavelength plate. (b) The probability distribution of the QW in resonator arrays with the walkers starting from $x = 0$ and the coin state chosen to be $(|H \rangle + |V \rangle) / \sqrt{2}$ for the first 6 steps. (c) The standard deviation of the QW (blue solid line), the classical random walks (the red solid line) and the QW in the resonator arrays (triangle markers) for the first 15 steps. The parameters are the same as that in Fig. \ref{PBS}.}
    \label{QW}
\end{figure}

With the tunable polarization beam splitter as the vector optomechanical system functions in hand, one can design the QW scheme in whispering-gallery-mode resonator arrays with reasonable arrangement. One of the straightforward ways to construct QW in resonator arrays is using the polarization states of the photon as coin states to determine which side of the cavity it will go into for the next step. Fig. \ref{QW}(a) shows an alternating scheme to implement QW which is governed by the operator $U = SC(\theta(x))$, with $S = \sum_{x}(|x \rangle \langle x+1| \bigotimes |0 \rangle \langle 0| + |x \rangle \langle x-1| \bigotimes |1 \rangle \langle 1|)$ the conditional transition operator and the position-dependent coin operator can be expressed as 

\begin{align}
    C(\theta(x)) &= I_x \bigotimes P(\theta(x)),\nonumber\\ 
    I_x &= \sum_x |x \rangle \langle x|, \label{equation 16}\\ 
    P(\theta(x)) &= 
    \left(
    \begin{array}{cc}
        cos(\theta(x)) & sin(\theta(x))\\
        sin(\theta(x)) & -cos(\theta(x))
    \end{array}
    \right). \nonumber 
\end{align}

$x$ is the position of the walker and $ \{ |0 \rangle, |1 \rangle \} $ are the two orthogonal coin states corresponding to the vertical and horizontal polarization of the photons, respectively. $I_x$ is the identity operator. $P(\theta(x))$ indicates there is a rotation for the coin states after every step and can be realized by half wavelength plate in our scheme. $\theta(x)$ depicts the rotation angle of the half wavelength plate dependent on the position of the walker for the generality. 

For the parameters in Fig. \ref{PBS}, the transmission of $T_{2\updownarrow}$ is $90\%$ and the transmission of $T_{4\leftrightarrow}$ is $100\%$ when $\theta = 0$. Considering the loss of the system, there are position and polarization dependent loss operator $L$ after the condition operator $S$ and the coin operator $C$ in each step which can be shown as

\begin{align}
    L = I_x \bigotimes
    \left(
    \begin{array}{cc}
        l_{1x} & 0\\
        0 & l_{2x}
    \end{array}
    \right),  0 \le l_{1x}, l_{2x} \le 1. \label{equation 16}
\end{align}

Fig. \ref{QW}(b) and Fig. \ref{QW}(c) show the probability distribution of the first 6 steps and the standard deviation of the first 15 steps of QW in the passive resonator arrays. Unlike classical random walks, the probability of the edge position is much higher than the probability of $x = 0$ for the QW case. It is inferred that the behavior of the walks in the resonator arrays matches with the quantum case. The most important difference between the QW and the classical random walks is that the standard deviation of the QW is proportional to the number of the steps $s$ while the classical random walks is proportional to $\sqrt{s}$. To make it clear, Fig. \ref{QW}(c) demonstrates the standard deviation of the quantum case (the blue solid line), the classical case (the red solid line) and the case in resonator arrays (the triangle markers). It can be found that the behavior of this case is similar to the quantum case, which solids the walks in our case is indeed the QW. Notice that the standard deviation of the first few steps close to the classical case because of the loss of the vector system. 

The results in Fig. \ref{kappa} show that the loss of the photons with different polarization is adjustable in the vector system, which provides an alternating way to implement PT-symmetric QW \cite{xiao2017observation} in the resonator arrays. Implementation of the PT-symmetric discrete-time QW allows us to observe different topological phases and have potential value in designing topological device taking advantages of the robustness of these phases to a variety of perturbations including impurities, decoherence, interactions, and explicit breaking of symmetries.

\section{OPTOMECHANICALLY INDUCED FARADAY EFFECT \label{faraday effect}}

\begin{figure}
    \centering
    \includegraphics[width=\linewidth]{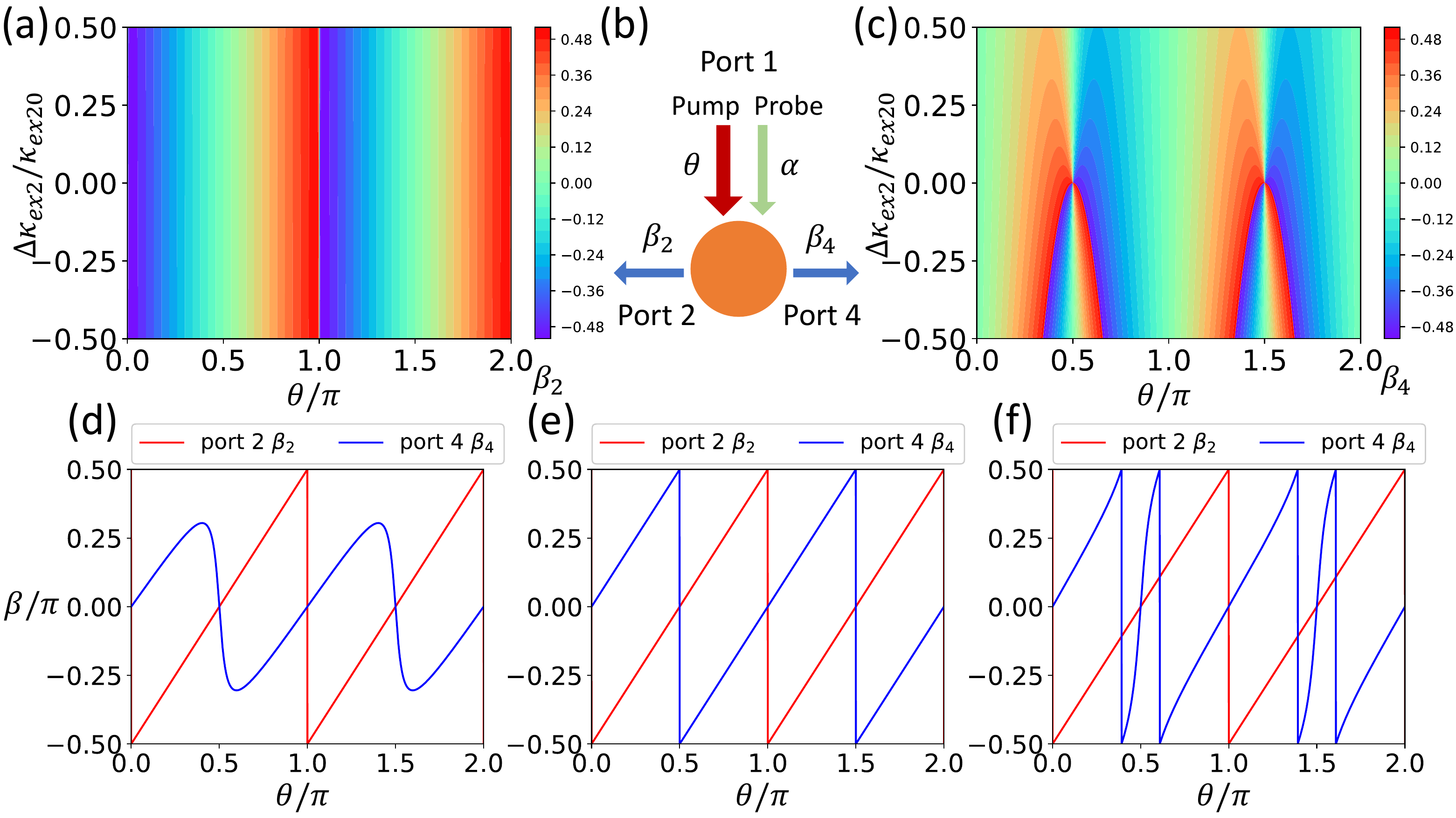}
    \caption{The angle between the polarization of the output field and the horizontal mode $\beta / \pi$ as a function of $\theta/\pi$ and $\Delta \kappa_{ex2} / \kappa_{ex20}$: (a) port 2 and (c) port 4. (b) Schematic of the input and output field polarization angles in the vector optomechanical system. The angle $\beta / \pi$ of different ports as a function of $\theta$ with different value of $\Delta \kappa_{ex2} / \kappa_{ex20}$: (d) 0.25, (e) 0, and (f) -0.25. $\kappa_{ex20}/2\pi = 8$ MHz. The other parameters are the same as that in Fig. \ref{PBS}.}
    \label{faraday}
\end{figure}

As mentioned above, there is OMIFE in the vector system. Fig. \ref{faraday} (b) depicts the schematic of the input and output field polarization angles in the model. In Fig. \ref{faraday}, we fix the polarization angle of the probe field at $\pi / 4$. Note the polarization angle of port 2 and port 4 are $\beta_2$ and $\beta_4$, which satisfy

\begin{align}
    tan(\beta_2) &= \frac{a_{\updownarrow}}{a_{\leftrightarrow}}, \label{equation 14}\\
    tan(\beta_4) &= \frac{\epsilon_{r\updownarrow} - \sqrt{\kappa_{ex1}} a_{\updownarrow}}{\epsilon_{r\leftrightarrow} - \sqrt{\kappa_{ex1}} a_{\leftrightarrow}}. \label{equation 15}
\end{align}

The parameter values in Fig. \ref{PBS} indicate the coupling between the cavity and the fiber is in the critical couple regime for both the optical modes as $\kappa_{ex1} = \kappa_{ex2} + \kappa_{0}$. It is necessary to explore the behaviors of the polarization angles in the under couple regime and in the over couple regime. Fig. \ref{faraday} (a) and Fig. \ref{faraday} (c) demonstrate the polarization angles of output field of port 2 and port 4 as a function of $\theta / \pi$ and $\Delta \kappa_{ex2} / \kappa_{ex20}$. In the $\Delta \kappa_{ex2} / \kappa_{ex20}$ axis we pick three values, i.e. 0.25, 0, and -0.25, which are in the under couple, critical couple, and over couple regime, respectively. It can be found that the behavior of the polarization angle of the output field of port 2 is maintained when changing the value of $\Delta \kappa_{ex2} / \kappa_{ex20}$ while the behavior of $\beta_4$ varies in the three coupling regimes. Furthermore, the corresponding polarization angle behaviors are shown in Fig. \ref{faraday} (d) - Fig. \ref{faraday} (f). The polarization of the output field of port 2 is always perpendicular to the polarization of the pump field. The reason is one can always construct a pair of new modes ($a_{//}$ and $a_{\bot}$) whose polarization are parallel to and perpendicular to the polarization of the pump field with $a_{\updownarrow}$ and $a_{\leftrightarrow}$. Due to OMIT effect the polarization of the output field of port 4 is parallel to the polarization of the pump field while the polarization of the output field of port 2 is parallel to the polarization of the pump field in the critical couple regime. In the under couple and over couple regimes the transmission of $a_{//}$ of port 4 is always 1 while the transmission of $a_{\bot}$ is not 0 anymore in the case of resonance. So the behaviors in the two regimes of $\beta_4$ is different from the critical couple regime. For port 2 the amplitude of transmission of $a_{\bot}$ will not change the polarization of the output field of port 2 as the transmission of $a_{//}$ is always zero in the three regimes under the condition of $\delta = \omega_m$. Due to Faraday effect in the vector optomechanical system, the polarization angles of port 2 and port 4 can be adjusted rapidly by tuning the polarization angle of the pump field and the coupling between the cavity and the fiber.

\section{CONCLUSION \label{conclusion}}

We have demonstrated a concise vector optomechanical system, consisting of two degenerate optical modes coupling with the same mechanical mode, which is a promising platform for continuously and all-optical tuning PBS. By changing the polarization angle of the pump field, one can control the polarization angles of different output ports. Furthermore, we study the OMIFE in this system and the different behaviors of polarization angles of the output field in the under couple, critical couple, and over couple regimes. Considering the feasibility of experiments we choose the values of the parameters from the previous experimental literature \cite{shen2018reconfigurable}. In this paper, we only focus on the case both the pump field and the probe field are linearly polarized. The input fields with circular polarization may lead to other interesting phenomenons which give more opportunities to complex polarization manipulations. As a significant application of the tunable PBS, an optional scheme of implementing QW in resonator arrays without the aid of other auxiliary systems is proposed in this paper. Furthermore, taking advantage of the tunable loss of the PBS, one can design novel QW platforms to detect and observe topological phases with a reasonable arrangement of the passive resonators. Our results prove the optomechanical system is a potential platform to manipulate the polarization states of the output fields and boost the process of applications of the optomechanical system.

\begin{acknowledgments}

We are grateful to Yuhao Qin for useful discussion and generous help about the code. This work is supported by the National Natural Science Foundation of China (61727801, 12025401 and U1930402); National Key Research and Development Program of China (2017YFA0303700); The Key  Research and Development Program of Guangdong province  (2018B030325002); Beijing Advanced Innovation Center for Future Chip (ICFC); Tsinghua University Initiative Scientific Research Program.

\end{acknowledgments}

%



\end{document}